\documentclass[reprint,amsmath,amssymb,aip,]{revtex4-1}
\usepackage{amsmath}
\usepackage{braket}
\usepackage{amsfonts}
\usepackage{graphicx}
\usepackage{dcolumn}
\usepackage{url}
\usepackage{bm}
\usepackage[colorlinks=true,citecolor=blue,linkcolor=blue,urlcolor=blue]{hyperref}
\usepackage{setspace}

\begin{document}

\title{The effect of Non-centrosymmetricity on optical and electronic properties of cubic BaHfO$_{3}$ perovskite structure}

\author{Tewodros Eyob}
\email{tewodros.eyob@aau.edu.et}
\author{Kenate Nemera}
\email{kenate.nemera@gmail.com}
\author{Lemi Demeyu}
\email{lemi.demeyu@aau.edu.et}
\affiliation{Department of
Physics, Addis Ababa University, P.O. Box 1176, Addis Ababa, Ethiopia}
\date{\today}
\begin{abstract}
The effect of Non-centrosymmetricity on electronic and optical properties has been investigated using grid based projector augmented wave method code GPAW. The calculation of band gap using GLLB-SC which improves on the exchange potential and explicitly estimates the derivative discontinuity is similar to experimental result. Ferro-electricity of BaHfO$_{3}$ is increased as lattice distortion kept on increasing. Static dielectric constant, index of refraction only show slight change and in good agreement with experimental and other results. Absorption reaches highest peaks in visible frequency range, and absorption coefficient decreased as light goes through BaHfO$_{3}$ bulk, this in turn enhance excitonic property. Exciton binding energy  increased over lattice distortion sites. The presented results are important  in connection with development of typical  perovskite optoelectronic device properties. 
\end{abstract}

\maketitle


\section{INTRODUCTION}

Perovskite based materials have attracted much theoretical and experimental attention in recent years. Primarily, alkalne metal hafnate are of practical importance in numerous applications. These materials are best-known
for their adaptability to exhibit several unique properties due to unstablity to lower energy of structural distortions\cite{1}.

\begin{figure}[htp!]
\centering
\includegraphics[width= 3.0in]{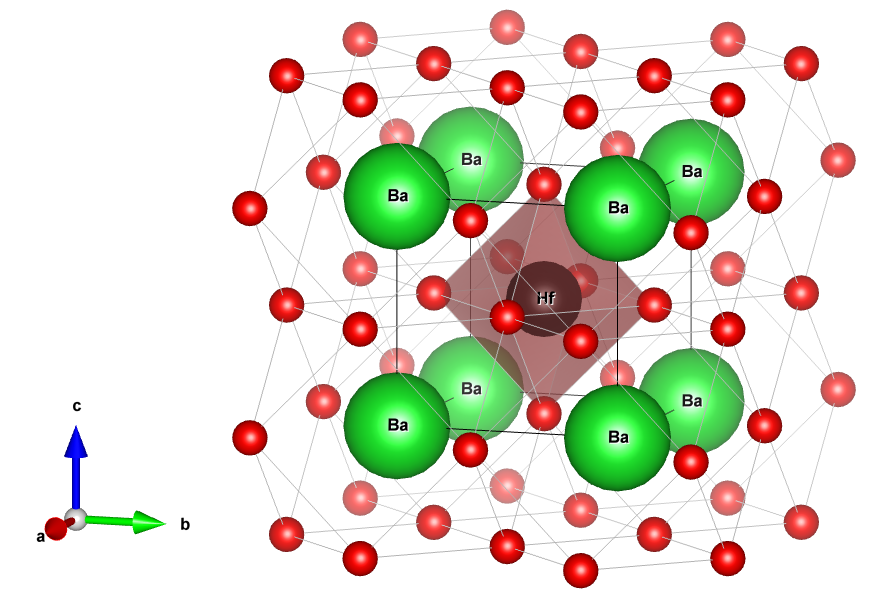}
\caption{\label{akqi} Crystal structure of Cubic BaHfO$_{3}$ perovskits with octahedron of O$_{6}$ surrounding Hf atom.}
\end{figure}

Cubic BaHfO$_{3}$ considered as ABO$_{3}$ perovskite oxides, where an A atom positioned at (0,0,0) corner of cube , a B atom  at body center position(1/2, 1/2, 1/2), and three O atoms at face centered positions(1/2, 1/2, 0) forming a BO$_{6}$ regular octahedron\cite{2}(see Fig.\ref{akqi}).
Several works have shown that the cubic phases of Pnma 
perovskites are unstable to both ferroelectric and octahedral
rotation distortions\cite{3,4,5,6}.
In present work we follow different  a approach, and chose unique
lattice displacement as distortions site in range of (3.85, 4.5)
for slight change of centrosymmetricity, and studied the effect 
on electronic, excitonic, ferroelectric and optical properties of 
BaHfO$_{3}$.
\section{COMPUTATIONAL METHOD}
We have implemented the Perdew-Burke-Ernzerhof(PBE) and Gritsenko, van Leeuwen, van Lenthe and Baerends potential(GLLB-SC) potentials to the grid-based projector augmented wave method code GPAW\cite{7,8} for Bandstructure calculations. It is a pseudopotential free approach, which allows more accurate and controlled description of electronic structure than the conventional pseudopotential approximations. For PAW core electrons the frozen-core approximation is used\cite{9,10}. Calculations were carried out self-consistently using GPAW PBE optimized lattice constant a=4.225$\AA$, Energy(E)=-36.147 eV, Bulk modulus(B)=157.939GPa(experimentally, a=4.171$\AA$), and  a 8x8x8 $\bf{k}$-points mesh over the entire BZ has been used.
Each material is calculated in the framework of Density Functional Theory\cite{11} using the GPAW code. All the proposed distortion sites are fully relaxed using the (PBE) functional\cite{12}.
\section{RESULTS AND DISCUSSION}
\begin{figure}[htp!]
\centering
\includegraphics[width= 3.50in]{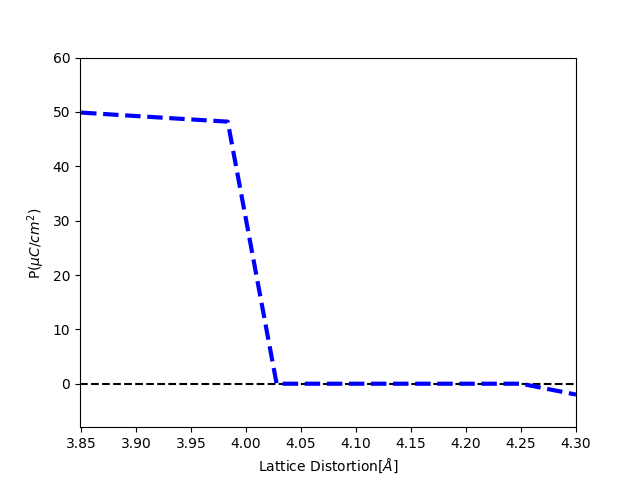}
\caption{\label{amj} Berry phase calculation of polarization($\mu/cm^{2}$) with respect to lattice distortion($\AA$).}
\end{figure}

We considered lattice displacement of 0.05 increase in range of (3.85, 4.5) as distortion site for the slight shift of centrosymmetricity. From Fig.\ref{amj} it is clearly seen that highest polarization obtained at start, and gradually decreases till it comes to zero where  barium hafnate's symmetry attained. Afterwards slow development of antiferroic property shown, this is because Hf atom, d-orbital from CB and p-orbital from VB contribution to Density of State(DOS) and non-centrosymmetricity(lattice distortion) allows VB to have higher electron than CB and result in polarization.In second part, symmetry proportionates electron from VB to CB. Thus, no net electron spin state for polarization of perovskites. In similar manner, the first part of Fig.\ref{amj} Hf atom, only p-orbital was contributing to VB, but in third part both p-orbital and s-orbital contributing to same band i.e CB. Therefore, Hf atom highly acted in reduction of net electron transfer from VB to CB, and resulted in almost zero ferroic property. However, at last Hf atom d-orbital and Ba, s-orbital contribution from CB is slightly greater than electron from VB. Hence, antiferroic property slowly grew up as distortion continued.

Here we presented representative sites and  calculated  the tolerance factor (t) which measures a geometric distortions of Peroviskets\cite{14}.
\begin{equation}
t=\frac{R_{A-O}}{\sqrt{2}R_{B-O}}
\end{equation}
where R$_{A-O}$ and R$_{B-O}$ are the ideal $A-O$ and $B-O$ bond lengths for a particular ABO$_{3}$ material calculated using the bond valence model. Geometry relaxed for all representative lattice sites until the maximum ionic force less than 0.05$(eV/$\AA$)$. Table.\ref{table:nonlin} shows calculated bond length between Ba-O and  Hf-O that  determines symmetry of BaHfO$_{3}$. At 3.85 site Hf-O bond pulled downward because of stronger force acted to shorten bond length and Ba-O bond pull upward in similar fashion, but at 4.171 and 4.2055 seems forces are in equilibrium, again Ba-O bond pulled upward, this chaotic disturbance introduce spontaneous polarization. Spontanous polarization is an important property of ferroelectricity of materials, and obtained when crystal alters centrosymmetricity\cite{15}.Here we used BerryPhase Phase-Space Approach implemented in GPAW\cite{16} for the calculation of spontaneous polarization for representative sites shown in Fig.\ref{amj}.
\begin{table}[h!]
\caption{Calculated minimum energy, Ba-O $\&$ Hf-O bond length and tolerance factor of representative lattice sites.} 
\centering 
\begin{tabular}{c c c c c c} 
\hline\hline 
\\
Sites($\AA$) & Energy(eV) & Ba-O($\AA$)& Hf-O($\AA$)& Tolerance(t)\\ [0.9ex] 
\hline 
3.85 & -35.17 & 4.69 & 1.95 & 1.70068 \\ 
4.0277& -28.12 & 3.11 & 4.52 & 0.48653 \\
4.171 & -35.94 & 2.95 & 2.08 & 1.00287 \\
4.2055 & -36.07 & 2.96 & 2.10 & 0.99668 \\
4.25 & -36.06 & 5.17 & 2.12 & 1.72441 \\ [1ex] 
\hline\hline 
\end{tabular}
\label{table:nonlin} 
\end{table}

\subsection{Electronic Properties}
\begin{figure}[htp!]
\centering
\includegraphics[width=3.5in]{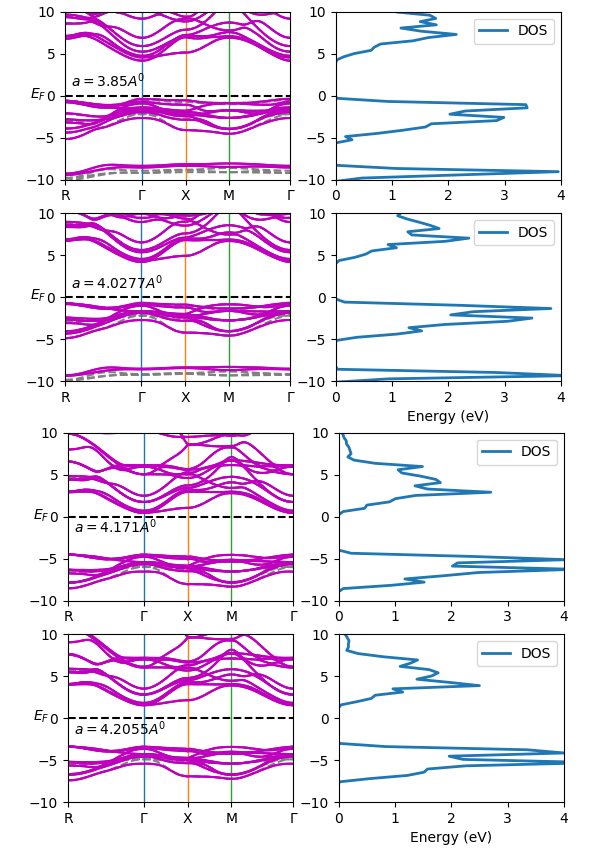}
\caption{\label{akq} Band structure and Fermi level change on lattice site 3,85, 4.171, 4.0277 and 4.2055 to their respective density of state.}
\end{figure}

We investigated the electronic structure and respective density of state compared among selected sites. Our calculated band gap is about 4 eV, which is in good agreement to other theoretical results(3.9eV)\cite{17}. As shown in Fig.\ref{akq}. Though fermi level oscillates in between band gap range, it does not not affect the band gap size. However, distortion strongly affects the crystal symmetry and electronic structure of the cubic peroviskite by stressing atomic orbitals, and confines mobility of electrons in either CB or VB, this situation leads to exhibit different electronic properties i.e ferroic, antiferroic, non-magnetic, P-type or N-type semiconductor so on. For band structure calculation we used an improved description of the GLLB-SC which improves on the exchange potential and explicitly estimates the derivative discontinuity\cite{18,19}. The GLLB-SC potential has been tested for cubic perovskites oxide systems with respect to experimental and G0W0 results giving an error below 0.5 eV. For all selected sites 3.85, 4.0277, 4.171, 4.2055 and respective fundamental gap in consecutive order is 6.164, 6.478, 6.55, 6.49 which  only differ from experimental result by 0.5 eV.
\subsection{Optical Properties}
The optical properties can be gained from the complex dielectric function which is mainly connected with the electronic structure.
\begin{equation}
\epsilon(\omega)=Re\epsilon(\omega)+Im\epsilon(\omega)
\end{equation}
 The imaginary part Im$\epsilon(\omega)$ is calculated from the electronic structure through the joint density of states\cite{20} and the momentum matrix elements between the occupied and the unoccupied wave functions within the selection rules and is given by
\begin{equation}
\epsilon(\omega)=\frac{2e^{2}\pi}{\Omega \epsilon_{0}} \sum|\bra{\psi_{k}^{c}}\hat{u}r\ket{\psi_{k}^{v}}|^{2}\delta(E_{k}^{c}-E_{k}^{v}-E) 
\end{equation}
where e is the electronic charge, and $\psi_{k}^{c}$ and $\psi_{k}^{v}$ are the conduction band (CB) and valence band (VB) wave functions at k, respectively.

\begin{figure}[htp!]
\centering
\includegraphics[width= 3.5in]{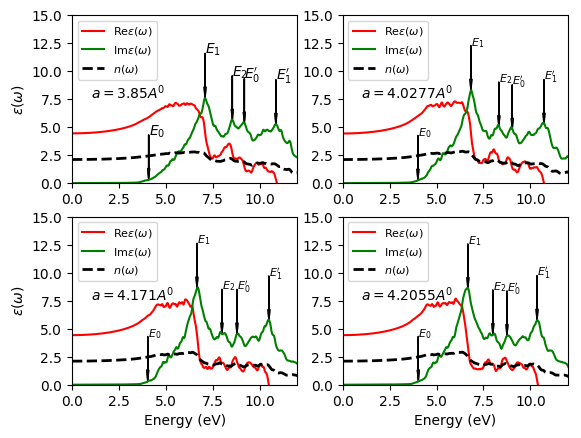}
\caption{\label{aknq}The calculated real part  Re$\epsilon(\omega)$ 
 and imaginary part Im$\epsilon(\omega)$ of the dielectric  $\epsilon(\omega)$ of BaHfO$_3$ at selected distortion sites.}
\end{figure}

The real part Re$\epsilon(\omega)$ can be extracted from the Kramer-Kronig relationship\cite{21}. Other optical constants such as the refractive index $n(\omega)$, and absorption coefficient $I(\omega)$ can be computed from complex dielectric function $\epsilon(\omega)$\cite{22}.
\begin{equation}
n(\omega)=\frac{1}{\sqrt{2}}\left\{\left[Re\epsilon^{2}(\omega)+Im\epsilon^{2}(\omega)\right]^{1/2}+Re\epsilon^{2}(\omega)\right\}^{1/2}
\end{equation}
and
\begin{equation}
I(\omega)=\frac{1}{\sqrt{2}}\frac{2\pi}{\lambda}\left\{\left[Re\epsilon^{2}(\omega)+Im\epsilon^{2}(\omega)\right]^{1/2}-Re\epsilon^{2}(\omega)\right\}^{1/2}
\label{koqw}
\end{equation}
The refractive index $n(\omega)$ and the dielectric function $\epsilon(\omega)$ of BaHfO$_{3}$ are shown in Fig.\ref{aknq} and compared in selected sites.

The overall behavior of $n(\omega)$ and static dielectric constants are found to be the same for all distortion sites and are in good agreement with other predicted study of $n(\omega)$ and dielectric constant\cite{23}. However, absorption peaks increased slightly as distortion goes far from centrosymmteric point, and turns on excitonic effect. 

\begin{figure}[htp!]
\centering
\includegraphics[width= 3.50in]{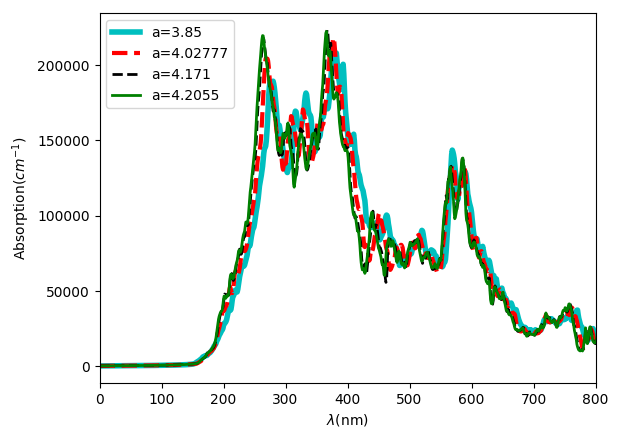}
\caption{\label{afgk}The absorption coefficient $I(\omega)$ for BaHfO$_3$ at selected distortion sites.}
\end{figure}

The absorption coefficient $I(\omega)$ described as the amount of light absorbed per thickness of BaHfO$_{3}$ and  calculated from  Eq.(\ref{koqw}). As illustrated in Fig.\ref{afgk} the  absorption  coefficient declines in the range of visible light(390-780nm) as distortion increases, and result in improved excitonic properties. 

The excitonic properties are very sensitive  to the grid of $\bf{k}$-points used because the contributions of Im$\epsilon(\omega, q\rightarrow 0)$ are very important. For the calculation of optical excitation we used  Bethe-Salpeter Equation(BSE), and included a number of valence and conduction bands. The imaginary part of the dielectric function, which is directly related to absorption spectrum, can be calculated by using the expression.
\begin{equation}
Im\epsilon(\omega)^{BSE}=\frac{16\pi^{2}e^{2}}{\omega^{2}}\sum|e.\bra{0}|v\ket{s}|^{2}\sigma(\omega-\Omega^{s})
\end{equation}

The term $|e.\bra{0}|v\ket{s}$ is called the velocity matrix element in the direction of the polarization of the light e\cite{24}.

For the calculation of dielectric function 4 unoccupied bands have been proven sufficient in the energy range under consideration. A simple scissor shift of 0.8 eV is used to match the onset of the absorption to experiment, neglecting local field effects and the electron-hole correlation has been considered.

\begin{figure}[htp!]
\centering
\includegraphics[width= 3.50in]{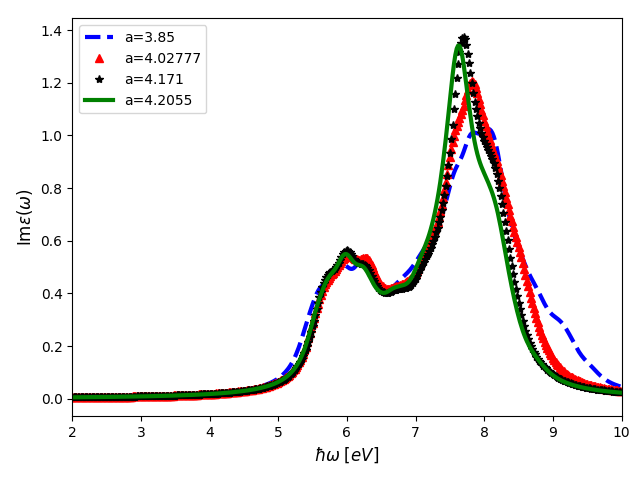}
\caption{\label{am} The dielectric function of BaHfO$_{3}$ obtained from the BSE, without local field effect.}
\end{figure}

BaHfO$_3$ is known by a large fundamental gap which is obtained from GLLBSC calculation as 6.164, 6.478, 6.55, 6.49 in order of lattice distortion 3.85, 4.0277, 4.171 and 4.25. Note that the optical spectrum at Fig.\ref{am}, the first excitonic peaks are at an energy of 6.236 eV, 6.0226 eV, 6.0016 eV and 6.00 eV, and the optical gap of BaHfO$_3$ in respective order of distortion sites are 3.85, 4.0277, 4.171 and 4.25, and also the G0W0 QP band gap are 4.01 eV, 4.2018 eV, 4.2779 eV, and 4.23 eV. Similarly, the excitonic binding energy are 1.99 eV, 1.8 eV, 1.72 eV, and 1.77 eV,  respectively. We observe from Fig.\ref{am} that as the stronger BaHfO$_{3}$ absorption is the lower dielectric function resulting in reduced screen effect and higher  exciton's binding energy and the longer lifetime. Therefore, reduced screening effect is because of lattice distortion sites.    

\section{SUMMARY}
In this work, We have computed using Density Functional Theory, spontaneous polarization, electronic, and optical properties for representative  values of lattice distortion. The calculated spontaneous polarization, electronic and optical spectra are in good agreement with experimental results. Efficiency of optoelectronic devices made of BaHfO{$_{3}$} material should consider distortion effect on exciton binding energy as per increase in absorption spectrum and decrease in absorption coefficient.Ferroic property of B-sites of ABO$_{3}$ system is highly affected by tolerance factor which measures distortion of perovskites. 

\begin{acknowledgments}
Addis Ababa University Physics Department and International Science Program, Uppsala University, Sweden are gratefully acknowledged for providing computational facilities.The authors are grateful to Prof.P.Singh, Addis Ababa University for helpful discussion.
\end{acknowledgments}

\bibliographystyle{aps-nameyear}      

\end{document}